\documentclass[draft,twocolumn,prb,amssymb,showpacs]{revtex4}
\input epsf

\begin{document}
%\draft
\title{Role of doped layers in dephasing of 2D electrons in quantum
well structures}
\author{G.~M.~Minkov}
\email{Grigori.Minkov@usu.ru}
\author{A.~V.~Germanenko}
\author{O.~E.~Rut}
\author{A.~A.~Sherstobitov}
\affiliation{Institute of Physics and Applied Mathematics, Ural
State University,  620083 Ekaterinburg, Russia}
\author{B.~N.~Zvonkov, E.~A.~Uskova, A.~A.~Birukov}
\affiliation{Physical-Technical Research Institute, University of
Nizhni Novgorod, 603600 Nizhni Novgorod, Russia}

\date{\today}
\begin{abstract}
The temperature and gate voltage dependences of the phase breaking
time are studied experimentally in GaAs/InGaAs heterostructures
with single quantum well. It is shown that appearance of states at
the Fermi energy in the doped layers leads to a significant
decrease of the phase breaking time of the carriers in quantum
well and to saturation of the phase breaking time at low
temperature.
\end{abstract}

 \maketitle

Inelasticity of electron-electron interaction is the main
mechanism of dephasing of the electron wave function in
low-dimensional systems at low temperature. Whereas the theory
predicts divergence of the phase breaking time $\tau_\varphi$
with decreasing temperature,\cite{Tf} an unexpected saturation of
$\tau_\varphi$ at low temperatures has been experimentally found
in one- and two-dimensional structures.\cite{r1,r2} These
observations rekindle a particular interest to dephasing.

In order to perform transport experiments, a $\delta$- or
modulation doped layer is arranged in semiconductor
heterostructures. Usually the doped layer is spaced from the
quantum well, all carriers leave the impurities and pass into the
quantum well. In some cases a fraction of the carriers remain in
the doped layer. As a rule, these carriers do not contribute to
the DC conductivity because they are localized in fluctuations of
long range potential. In other words, the percolation threshold of
doped layer is above the Fermi level. In this paper we
demonstrate that presence of the carriers and/or empty states in
doped layer at the Fermi energy can contribute to dephasing and
its temperature dependence.

The main method of experimental determination of the phase
breaking time is an analysis of the low-field negative
magnetoresistance, resulting from destruction of the interference
quantum correction to the conductivity. We have measured the
negative magnetoresistance in a single-well gated heterostructures
GaAs/In$_{x}$Ga$_{1-x}$As. The heterostructures were grown by
metal-organic vapor-phase epitaxy on a semiinsulator GaAs
substrate.  They consist of $0.5$-mkm-thick undoped GaAs epilayer,
a Sn $\delta$ layer, a 60-\AA\ spacer of undoped GaAs, a 80-\AA\
In$_{0.2}$Ga$_{0.8}$As well, a 60-\AA\ spacer  of undoped GaAs, a
Sn $\delta$ layer, and a 3000-\AA\ cap layer of undoped GaAs. The
samples were mesa etched into standard Hall bridges and then Al
gate electrode was deposited onto the cap layer by thermal
evaporation. The measurements were performed in the temperature
range $1.5-16$ K at magnetic field $B$ up to 6 T. The discrete at
low-field measurements was $5\times 10^{-5}$ T. The electron
density was found from the Hall effect and from the Shubnikov-de
Haas oscillations. These values coincide with an  accuracy of 5\%
over the entire gate-voltage range.

\begin{figure}
 \epsfclipon
 \epsfxsize=\linewidth
 \epsfbox{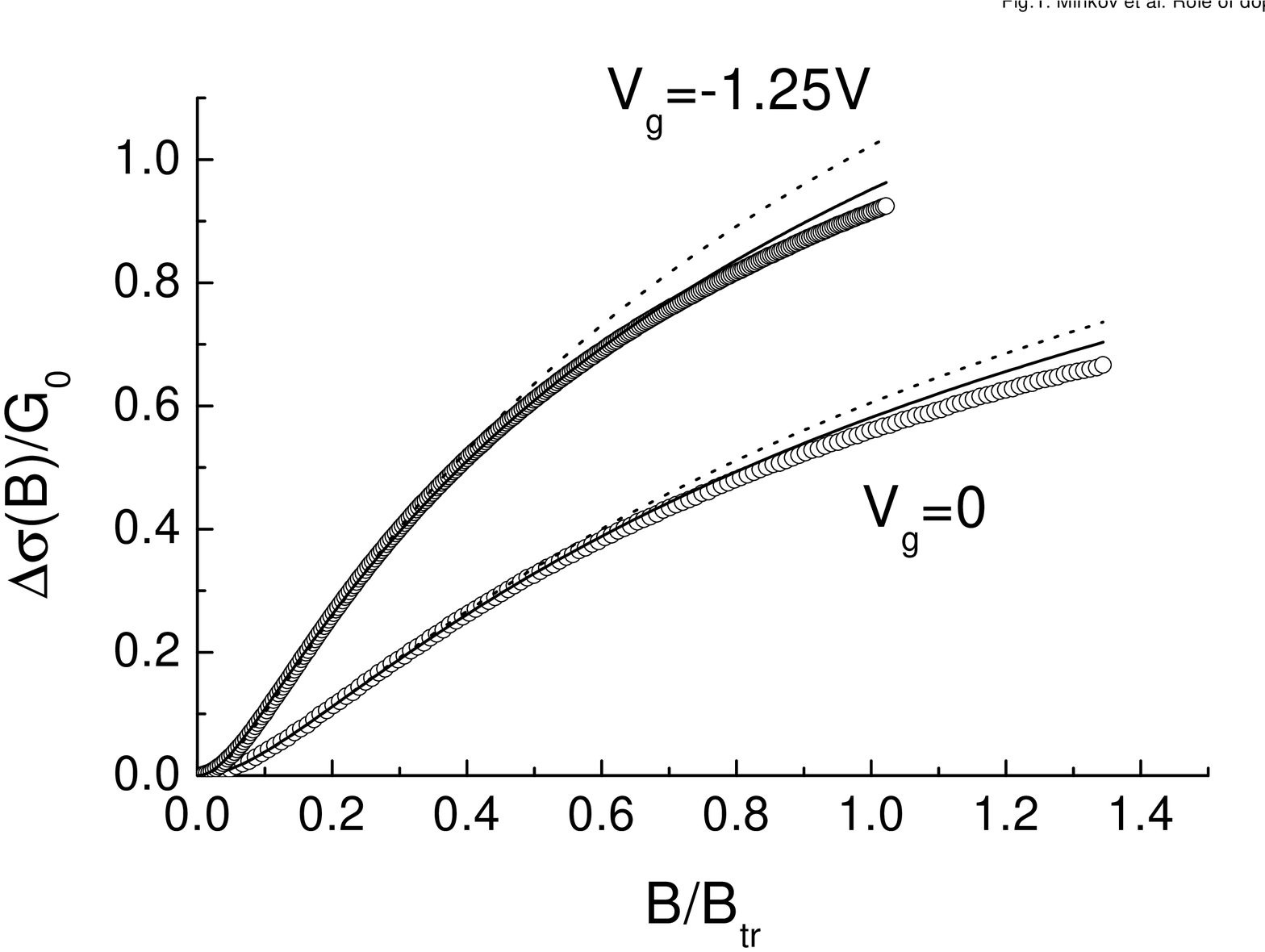}
 \caption{Conductivity changes versus magnetic field at T=4.2 K for two
gate voltages. Symbols are the experimental data. Curves are the
results of best fit by Eq.~(\ref{eq1}), carried out over the
ranges $0-0.25B_{tr}$ (dotted curves) and $0-0.5 B_{tr}$ (solid
curves). The fitting parameters for the curves from the top to
bottom are: $\alpha=0.8$, $\tau_\varphi=0.87\times 10^{-11}$ sec;
$\alpha=0.70$, $\tau_\varphi=1.0\times 10^{-11}$ sec;
$\alpha=0.8$, $\tau_\varphi=0.73\times 10^{-11}$ sec;
$\alpha=0.68$, $\tau_\varphi=0.77\times 10^{-11}$ sec.
$B_{tr}=5\times 10^{-3}$ T for $V_g=0$ V, $B_{tr}=1.5\times
10^{-2}$ T for $V_g=-1.25$ V } \label{fig2}
\end{figure}

Varying the gate voltage $V_g$ from $+1.0$ to $-2.5$ V we changed
the electron density in quantum well from $8\times 10^{11}$ to
$3.5\times 10^{11}$ cm$^{-2}$ and the conductivity $\sigma$ from
$2.3\times 10^{-3}$ to $1.5\times 10^{-4}$ Ohm$^{-1}$. The
low-field magnetoconductivity
$\Delta\sigma(B)=\rho_{xx}^{-1}(B)-\rho_{xx}^{-1}(0)$ for two gate
voltages is shown in Fig.~\ref{fig2}. To determine the phase
breaking time we have used the standard procedure of fitting of
the low-field magnetoconductivity to the Hikami expression
\cite{r3}

\begin{eqnarray}
\Delta\sigma(B) &=&\alpha G_0 \biggl[ \psi\left(\frac{1}{2}
+\frac{\tau_p}{\tau_\varphi}\frac{B_{tr}}{B}\right)- \nonumber \\
&-& \psi\left(\frac{1}{2}+\frac{B_{tr}}{B}\right)-
\ln{\left(\frac{\tau_p}{\tau_\varphi}\right)}\biggr], \label{eq1}
\end{eqnarray}
where $G_0=e^2/(2\pi^2 \hbar)$, $B_{tr}=\hbar/(2el^2)$, $l$ is the
mean-free path, $\tau_p$ is the momentum relaxation time,
$\psi(x)$ is the digamma function, and $\alpha$ is equal to unity.
This expression was obtained within the diffusion approximation.
Nevertheless, as shown in Ref.~\onlinecite{r4}, with $\alpha$ less
than unity it can be used for analysis of the experimental data
even beyond the diffusion approximation, giving the value of
$\tau_\varphi$ close to true one. The results of fitting carried
out in two magnetic field ranges are presented in Fig.~\ref{fig2}.
One can see that the values of fitting parameters somewhat depend
on the magnetic field range. However, the difference in
$\tau_\varphi$ does not exceed $20$\%. The value of $\alpha$ is
lower than unity and lies within the interval $\alpha=0.6-0.8$.
This is result of the fact that in the structure investigated the
ratio $\tau_p/\tau_\varphi$ is not low enough:\cite{r4}
$\tau_p/\tau_\varphi\simeq 0.014-0.1$.

\begin{figure}[t]
 \epsfclipon
 \epsfxsize=\linewidth
 \epsfbox{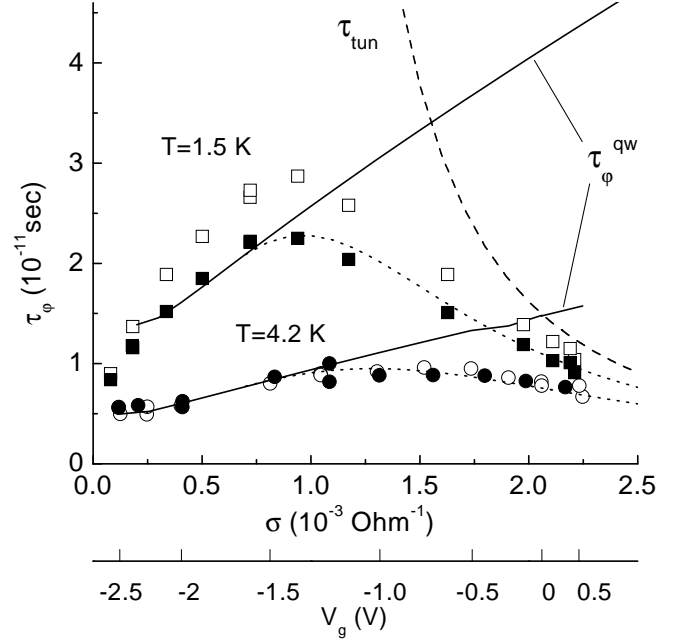}
 \caption{Conductivity (gate voltage) dependences of $\tau_\varphi$ obtained by
fitting of the experimental $\Delta\sigma$-versus-$B$ curves by
Eq.~(\ref{eq1}) in the magnetic field range $0-0.25 B_{tr}$ (open
symbols) and $0-0.5 B_{tr}$ (solid symbols) for two temperatures.
Solid lines are the theoretical dependences given by
Eq.~(\ref{eq2}). The $\tau_{tun}$-versus-$\sigma$ line is obtained
as described in text. Dotted lines are
$\tau_\varphi=\left((\tau_\varphi^{qw})^{-1}+(\tau_{tun})^{-1}\right)^{-1}$.}
\label{fig3}
\end{figure}

Figure~\ref{fig3} shows the conductivity dependence of
$\tau_\varphi$ for two temperatures.  One can see that
$\tau_\varphi$-versus-$\sigma$ plot exhibits maximum:
$\tau_\varphi$ increases with increasing conductivity while
$\sigma<10^{-3}$ Ohm$^{-1}$ and decreases at higher $\sigma$
values. Qualitatively this behavior is independent of the fitting
range. The maximum is more pronounced at lower temperature.

Nonmonotonic conductivity dependence of $\tau_\varphi$ is in
conflict with the theoretical prediction. In 2D systems the main
phase breaking mechanism at low temperature is inelasticity of the
electron-electron interaction and the phase breaking time has to
increase monotonically with $\sigma$:\cite{Tf}
\begin{equation}
\tau_\varphi^{th}=\frac{\hbar}{kT}\frac{\sigma}{2\pi
G_0}\left[\ln\left(\frac{\sigma}{2\pi G_0}\right)\right]^{-1}.
\label{eq2}
\end{equation}

\begin{figure}
\epsfclipon
 \epsfxsize=5cm
\epsfbox{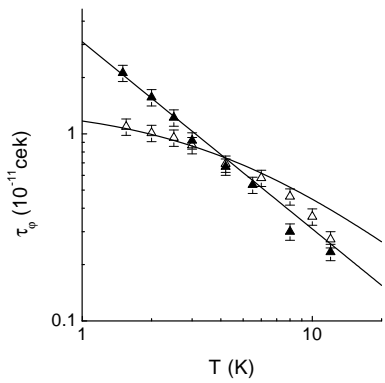}
  \caption{The temperature dependences of $\tau_\varphi$ at
$V_g=-1.8$~V, $\sigma=0.53\times 10^{-3}$ Ohm$^{-3}$ (solid
triangles) and $V_g=+0.5$~V, $\sigma=2.2\times 10^{-3}$ Ohm$^{-3}$
(open triangles). The lines are results of calculation described
in text with $\tau_{tun}=\infty$ for $V_g=-1.8$ V and
$\tau_{tun}=1.4\times 10^{-11}$~sec for $V_g=+0.5$~V.}
\label{fig5}
\end{figure}

As is seen from Fig.~\ref{fig3} the conductivity dependence of
$\tau_\varphi$ is close to the theoretical one only when $\sigma <
(1.0 - 1.2)\times 10^{-3}$~Ohm$^{-1}$, but significantly deviates
for higher  $\sigma$. In addition, the increase of $\sigma$ leads
to changing in the temperature dependence of $\tau_\varphi$ (see
Fig.~\ref{fig5}). When $\sigma=(0.2 - 1.2)\times
10^{-3}$~Ohm$^{-1}$, the temperature dependence of $\tau_\varphi$
is close to $T^{-1}$ predicted theoretically. At higher $\sigma$,
the $\tau_\varphi$-versus-$T$ plot shows the saturation at low
temperature.

\begin{figure}
\epsfclipon
 \epsfxsize=7cm
 \epsfbox{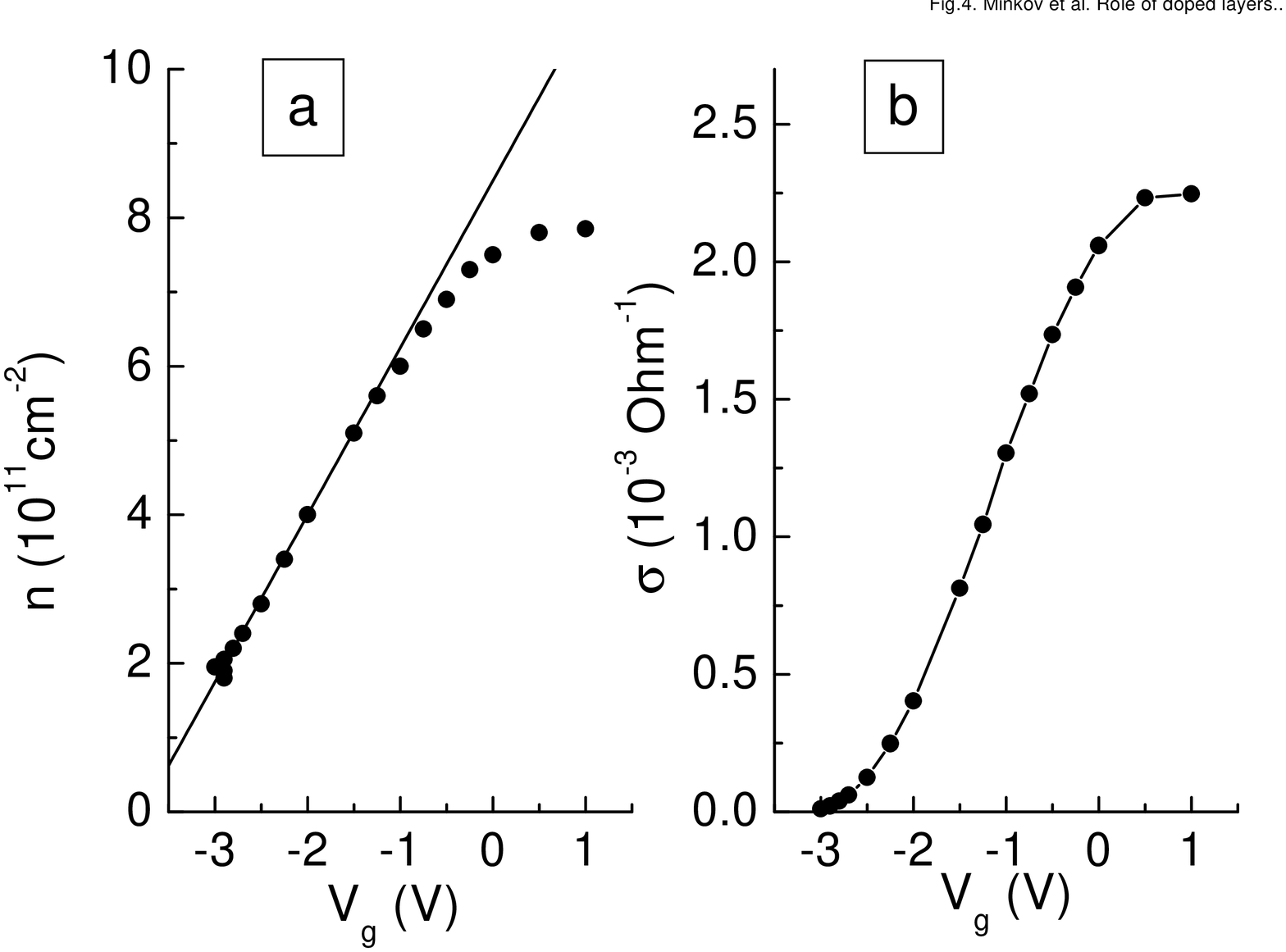}
\caption{The experimental gate voltage dependences of the electron
density in quantum well (a) and structure conductivity (b) at
T=4.2 K (symbols). The straight line in (a) is the calculated
total electron density with $n(0)=8.4\times 10^{11}$~cm$^{-2}$. }
 \label{fig1}
\end{figure}

To interpret the experimental temperature and conductivity
dependences of $\tau_\varphi$, let us first analyze the variation
of electron density in the quantum well $n$ with the gate voltage
[see Fig.~\ref{fig1}(a)]. The total electron density $n_t$ in a
gated structure has to be given by the simple expression
$n_t(V_g)=n(0)+V_g\, C/|e|$, where $C$ is the gate--2D channel
capacity per centimeter squared. Straight line in
Fig.~\ref{fig1}(a) represents this dependence obtained with $n(0)$
as fitting parameter and $C=\varepsilon/(4\pi d)$, where
$d=3000$\AA\ is the cap-layer thickness, $\varepsilon=12.5$. One
can see that over the range of $V_g$ from $-2.5$ to $-0.5$ V the
experimental data are close to the calculated dependence, whereas
at $V_g>-0.5$~V the electron density in the quantum well is less,
than the total density $n_t$.

\begin{figure}
\epsfclipon
 \epsfxsize=8cm
\epsfbox{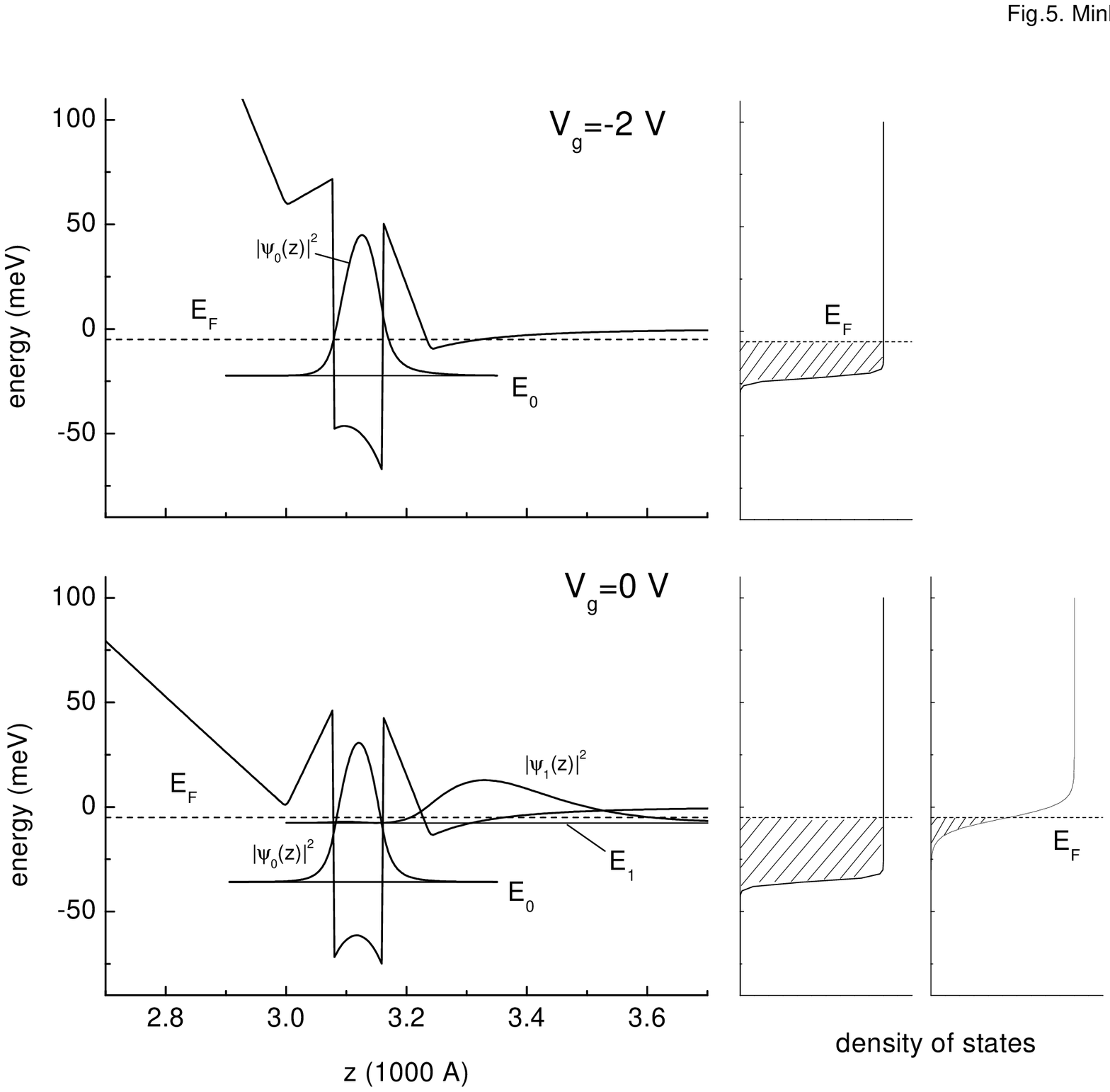} \caption{The calculated energy diagram (left
panels) and sketch for density of electron states (other panels)
for the structure investigated at two gate voltages: $V_g=-2.0$
and $0$~V. The energy levels of size quantization $E_i$ and
corresponding wave functions squared $|\Psi_i(z)|^2$ are presented
also. The distance $z$ is measured from the gate. The height of
the Schottky barrier was taken to be equal to $700$~meV
($V_g=0$).} \label{fig41}
\end{figure}

Such a behavior can be understood from inspection of
Fig.~\ref{fig41}, which presents the energy diagram of the
structure investigated for two gate voltages. Selfconsistent
calculation shows that at $V_g<-0.5$ V there are only the states
located in the quantum well (upper panels). At $V_g>-0.5$ V the
states located in $\delta$-layer appear (lower panels). However,
it should be borne in mind that the strong potential fluctuations
in $\delta$-layer leads to formation of the tail in density of
these states. At low electron density in $\delta$-layer, when the
Fermi level lies within the tail, the electrons are localized in
potential fluctuations and, thus, do not contribute to the
structure conductivity. The absence of both magnetic field
dependence  of the Hall coefficient and positive
magnetoresistance in wide range of magnetic field shows that this
situation occurs in our structures while $(n_t-n)<10^{11}$
cm$^{-2}$.

\begin{figure}
\epsfclipon
 \epsfxsize=7cm
\epsfbox{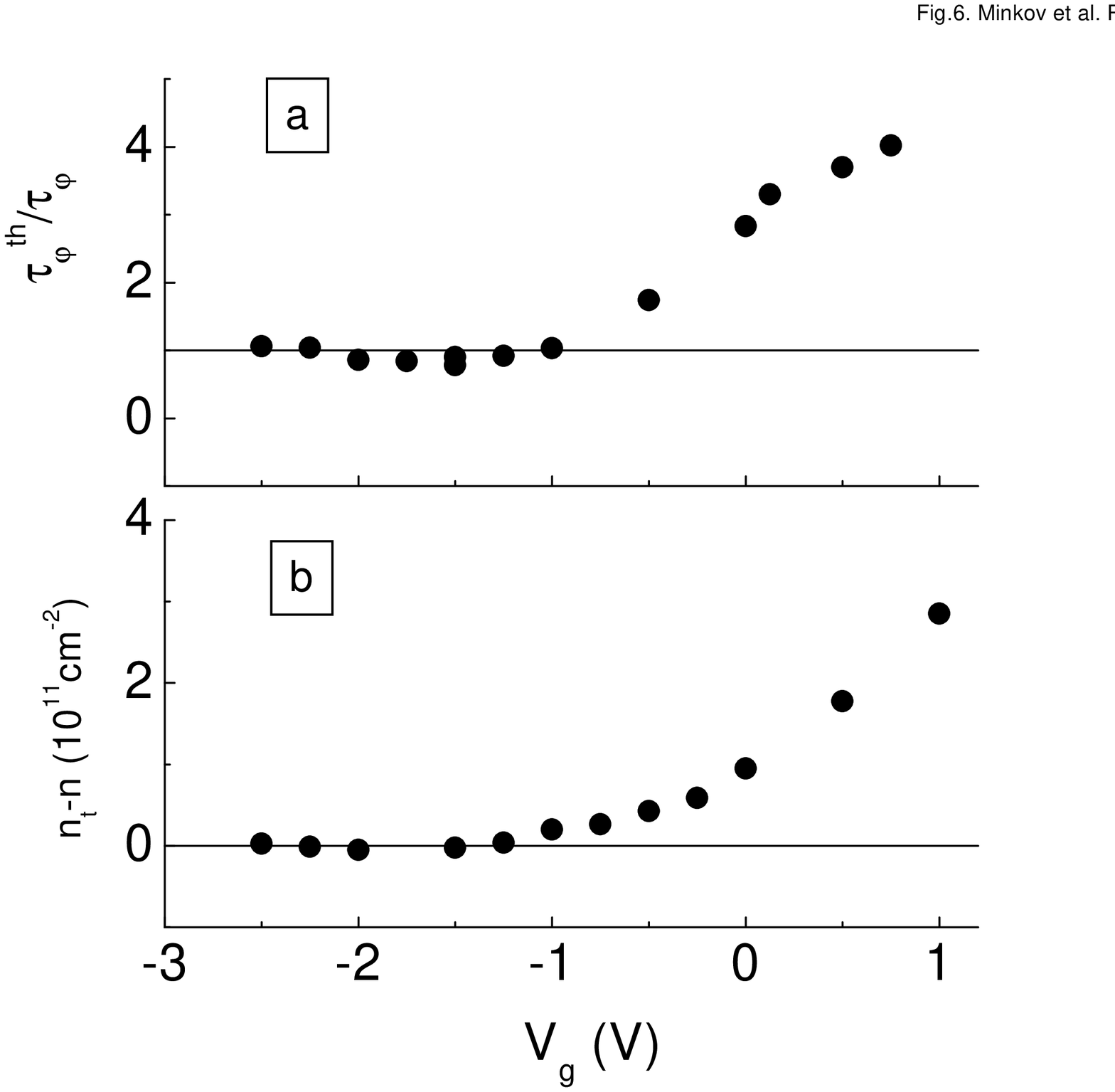}
 \caption{The gate voltage dependences of the ratio
$\tau_\varphi^{th}/\tau_\varphi$ for $T=1.5$ K (a) and difference
between the total electron density and density of electrons in
the quantum well (b).} \label{fig4}
\end{figure}

Now we are in position to put together the gate voltage
dependences of the phase breaking time  and those of the electron
density in $\delta$-layer. Figure~\ref{fig4} shows the ratio
$\tau_\varphi^{th}/\tau_\varphi$ as function of $V_g$, where
$\tau_\varphi^{th}(V_g)$ has been found from Eq.~(\ref{eq2}) with
the use of experimental $\sigma$-versus-$V_g$ dependence presented
in Fig.~\ref{fig1}(b). It is clearly seen that the experimental
values of $\tau_\varphi$ are close to theoretical ones, i.e.,
$\tau_\varphi^{th}/\tau_\varphi\simeq 1$, when there are no
electrons in $\delta$-layer ($V_g<-0.5$ V) and become to be
significantly less when carriers appear therein ($V_g>-0.5$ V).
Thus, an additional mechanism of phase breaking for electrons in
the quantum well arises when the $\delta$-layer is being
populated.

One of such mechanisms can be associated with tunneling. Indeed,
appearance of electrons in $\delta$-layer means arising of empty
states at the Fermi energy therein, and, as sequence, tunneling of
electrons between quantum well and $\delta$-layer. In this case,
an electron moving over closed paths (just they determine the
interference quantum correction) spend some time within the
$\delta$-layer. Due to low value of local conductivity in
$\delta$-layer, the electron fast losses the phase memory
therein.\footnote{Strictly speaking,  the phase breaking
mechanisms in doped layers, where electrons occupy the states in
the tail of density of states, are the subject of additional
study, but it seems no wonder that dephasing in these layers
occurs faster than in quantum well.} If the phase breaking time in
$\delta$-layer is much shorter than both the tunneling time
$\tau_{tun}$ and the phase breaking time in quantum well
$\tau_\varphi^{qw}$, the effective phase breaking time will be
given by simple expression
\begin{equation}
\frac{1}{\tau_\varphi}=\frac{1}{\tau_\varphi^{qw}}+\frac{1}{\tau_{tun}}.
\label{eq3}
\end{equation}

Let us analyze our experimental results from this point of view.
We can find the gate-voltage dependence of $\tau_{tun}$ and thus
the conductivity one from Eq.~(\ref{eq3}) using
$\tau_\varphi^{qw}$ calculated from Eq.~(\ref{eq2}) and the
experimental values of $\tau_\varphi$ for $T=1.5$ K. The results
presented in Fig.~\ref{fig3} by dashed line was obtained with
$\tau_\varphi$ determined from the fitting within the magnetic
field range $0-0.5B_{tr}$. It is seen that $\tau_{tun}$ sharply
decreases when the conductivity increases. Such a behavior is
transparent. The tunneling probability is proportional to the
density of final states. In our case the final states are the
states in the tail of the density of states of $\delta$-layer,
therefore their density at the Fermi energy increases when $V_g$
increases (see Fig.~\ref{fig41}).

Since the tunneling is temperature independent process,
Eq.~(\ref{eq3}) with the $\tau_{tun}$-versus-$V_g$ dependence
found above has to describe the experimental gate-voltage
dependence of $\tau_\varphi$ for any temperatures. Really, as is
seen from Fig.~\ref{fig3} the results for $T=4.2$ K are very
close to calculated curve.

The tunneling  has to change the temperature dependence of
$\tau_\varphi$ also. Because $\tau_\varphi^{qw}$ increases with
the temperature decrease as $T^{-1}$, the effective phase breaking
time has to saturate at low temperature. In Fig.~\ref{fig5} the
temperature dependences of $\tau_\varphi$ found from
Eq.~(\ref{eq3}) with $\tau_\varphi^{qw}$ calculated from
Eq.~(\ref{eq2}) and $\tau_{tun}$ determined above are presented.
Good agreement between the calculated and experimental results
shows that this model naturally describes the low-temperature
saturation of $\tau_\varphi$.

Thus, taking into consideration the electron tunneling between
quantum well and $\delta$-layer we have explained both the
gate-voltage and temperature dependences of phase breaking time.

Another possible mechanism of phase breaking for systems where
carriers occur not only in the quantum well is their interaction
with carriers in doped layer. Inelasticity of this interaction can
be of importance. High efficiency of such interaction can be
result of exciting of local plasmon modes in $\delta$-layers, for
example. We do not know the papers where inelasticity of the
interaction between electrons in quantum well and in doped layer
was taken into account.

In conclusion, the study of weak localization in gated structures
shows that appearance of carriers in doped layer leads to
decreasing of the phase breaking time and changing in the
conductivity and temperature dependences of $\tau_\varphi$. It
should be mentioned  that the $\delta$- or modulation doped layers
are arranged in all the heterostructures suitable for transport
measurements, to create the carriers in quantum well. The
mechanisms discussed above can be important for phase breaking,
even though the doped layers do not contribute to the
conductivity.

\subsection*{Acknowledgment}
This work was supported in part by the RFBR through Grants No.
00-02-16215, No. 01-02-06471, and No. 01-02-17003,  the Program
{\it University of Russia} through Grants No.~990409 and
No.~990425, the CRDF through Award No. REC-005, and the Russian
Program {\it Physics of Solid State Nanostructures}.


\begin{thebibliography}{}
\bibitem{Tf}
B. L. Altshuler and A. G. Aronov, in {\it Electron - Electron
Interaction in Disordered Systems}, edited by A. L. Efros and M.
Pollak, (North Holland, Amsterdam, 1985) p.1.

\bibitem{r1}
P.~Mohanty, E.~M.~Q.~Jarivala, and R.~A.~Webb, Phys. Rev. Lett.
{\bf 78}, 3366 (1997).

\bibitem {r2} W. Poirier, D. Mailly, and M. Sanquer,
preprint available at http://xxx.lanl.gov/abs/cond-mat/9706287.
 %(unpublished).

\bibitem{r3} S.~Hikami, A.~Larkin, and Y.~Nagaoka, Prog. Theor.
Phys. {\bf 63}, 707 (1980).

\bibitem{r4} G.~M.~Minkov, A.~V.~Germanenko,
V.~A.~Larionova, S.~A.~Negashev, and I.~V.~Gornyi, Phys. Rev. B
{\bf 61}, 13164 (2000).
\end{thebibliography}
\end{document}